\documentclass[twocolumn,showpacs,showkeys,aps]{revtex4}
\usepackage{graphicx,amsmath,amsfonts,amssymb}
\newcommand{\be}{\begin{equation}}
\newcommand{\ee}{\end{equation}}

\begin{document}
\title{Comment on ``Soliton ratchets induced by excitation of internal modes"}
\author{Niurka R.\ Quintero}
\email{niurka@euler.us.es} \affiliation{Departamento de F\'{\i}sica Aplicada I,
E.\ U.\ P., Universidad de Sevilla, Virgen de \'Africa 7, 41011 Sevilla, Spain}
\affiliation{Instituto Carlos I de F\'{\i}sica Te\'orica y Computacional,
Universidad de Granada, 18071 Granada, Spain}

\author{Bernardo S\'anchez-Rey}
%\email{bernardo@us.es}
\affiliation{Nonlinear Physics Group, University of
Seville, Spain}

\author{Jes\'us Casado-Pascual}
%\email{}
\affiliation{Universidad de Sevilla, F\'{\i}sica Te\'orica, Apartado de Correos 1065, 
Sevilla 41080, Spain}

\date{\today}

\begin{abstract}
Very recently Willis et al. [Phys. Rev. E {\bf 69}, 056612 (2004)] have
used a collective variable theory to explain the appearance of a nonzero
energy current in an ac driven, damped sine-Gordon equation. In this
comment, we prove rigorously that the time-averaged energy current in an
ac driven nonlinear Klein-Gordon system is strictly zero.
\end{abstract}
\pacs{05.45.Yv, 05.45.-a, 05.60.Cd, 63.20.Pw, 02.30.Jr}

\maketitle

Recently several papers have been published trying to understand soliton
ratchets (see, for example,
Refs.~\cite{Flach,Salerno,Niurka1,Costant,Niurka2,Willis} and for a recent
review  chapter 9 in Ref.~\cite{Kivshar}, pp. 343--364). This phenomenon is a
generalization of the ratchet effect \cite{Ratchet} to spatially extended
systems, and manifests as a unidirectional motion of a soliton induced by
zero-average forces. A paradigmatic example is the driven, damped nonlinear
Klein-Gordon equation:
\begin{equation}
\phi _{,tt}(x,t)-\phi _{,xx}(x,t)
=-U^{\prime}\left[\phi(x,t)\right]+f(t) -\beta \phi _{,t}(x,t),
\label{KG}
\end{equation}
where $g_{,z}=\partial g/\partial z$, $f(t)$ is a periodic field with
period $T$ and zero time-average [i.e., $1/T \int_{0}^{T} d t \,
f(t)=0$], $\beta>0$ is the dissipation parameter determining the
inverse relaxation time in the system, and $U^{\prime}(z)$ is the
derivative with respect to $z$ of the potential $U(z)$. In this comment,
we will assume that the potential $U(z)$ is periodic with period
$\lambda$, and presents minima at $z_{j}=z_0+j \lambda$, with $j\in
\mathbb{Z}$. The ac driven, damped sine-Gordon equation considered in
Ref.~\cite{Willis} is a particular case of this more general problem,
with $U(z)=1-\cos(z)$ and
\begin{equation}
\label{force}
f(t)=-\left[\epsilon_1 \cos (\omega t)+ \epsilon_2 \cos( 2 \omega
t+\theta)\right].
\end{equation}
To fully specify the mathematical problem, the
partial differential equation (\ref{KG}) must be amended by both initial
conditions for $\phi(x,0)$ and $\phi_{,t}(x,0)$, and boundary conditions
for $\lim_{x\rightarrow \pm \infty}\phi(x,t)$.  Several boundary
conditions can be imposed to have a well-posed boundary value
problem. For instance, in the absence of the periodic field $f(t)$, it
is possible to choose the fixed boundary conditions: $\lim_{x\rightarrow
+\infty} \phi(x,t)= z_{l}$ and $\lim_{x\rightarrow -\infty} \phi(x,t)=
z_{m}$. In the presence of $f(t)$, the fixed boundary conditions become
incompatible with Eq.~(\ref{KG}), and they are usually replaced by the
aperiodic boundary conditions:
\begin{eqnarray}
\label{BC1}
\lim_{x\rightarrow +\infty} \phi(x,t)&=&\lim_{x\rightarrow -\infty}
\phi(x,t)+\lambda Q,\\
\label{BC2}
\lim_{x\rightarrow +\infty} \phi_{,x}(x,t)&=&\lim_{x\rightarrow -\infty}
\phi_{,x}(x,t),
\end{eqnarray}
where $Q\in \mathbb{Z}$ is the so-called topological charge. The
discrete version of these periodic boundary conditions are also the most
used in the numerical solution of Eq.~(\ref{KG}) (see, for example,
Refs.~\cite{Flach} and \cite{Niurka2}).

It can be derived from the continuity equation that the energy current
density generated by the field $\phi(x,t)$ in the absence of damping and
external forcing is given by $j(x,t)=-\phi_{,t}(x,t) \phi_{,x}(x,t)$ and,
consequently, the energy current reads
\begin{equation}
 J(t) = - \int_{-\infty}^{+\infty} dx \;\phi_{,x}(x,t) \phi_{,t}(x,t).
\label{momentum}
\end{equation}
The time-averaged energy current $\langle J\rangle$ is defined as the
limit
\begin{equation}
 \langle J \rangle = \lim_{\tau\rightarrow +\infty} \frac{1}{\tau}
 \int_{0}^{\tau} d t\; J(t).
\label{momentum2}
\end{equation}
In Ref.~\cite{Flach}, it has been proved by symmetry considerations that
a {\em necessary} condition for the appearance of a non-vanishing
time-averaged energy current is that either the potential presents
broken spatial symmetry, or the field $f(t)$ violates the symmetry property
\begin{equation}
\label{conditions}
f\left(t+\frac{T}{2}\right)=-f(t),
\end{equation}
or both simultaneously.  Following this idea, a collective variable
approach has been developed in Ref.~\cite{Willis} for the ac driven,
damped sine-Gordon equation with a field of the form (\ref{force}) which
leads to a non-vanishing time-averaged energy current. The purpose of
this comment is to prove that the time-averaged energy current, $\langle
J \rangle$, of a driven, damped nonlinear Klein-Gordon equation of the
form (\ref{KG}) is necessarily zero and, consequently, the above
mentioned results in Ref.~\cite{Willis} must be erroneous.

To prove that $\langle J \rangle =0$, firstly we will obtain an ordinary
differential equation for the energy current $J(t)$. In order to do
that, we differentiate with respect to time Eq.~(\ref{momentum}),
resulting
\begin{equation}
\dot{J}(t)=-\int_{-\infty}^{+\infty} dx \left[ \phi_{,xt}(x,t)
\phi_{,t}(x,t)+\phi_{,x}(x,t) \phi_{,tt}(x,t) \right].
\end{equation}
By making use of Eq.~(\ref{KG}) in the second term on the right-hand
side of the above expression, it is straightforward to write it in the
form
\begin{eqnarray}
\label{eq1}
\dot{J}(t)&=&-\frac{1}{2}\left\{\lim_{x\rightarrow +\infty}
\left[\phi_{,t}(x,t)\right]^2-\lim_{x\rightarrow -\infty}
\left[\phi_{,t}(x,t)\right]^2\right\} \nonumber\\
&&-\frac{1}{2}\left\{\lim_{x\rightarrow +\infty}
\left[\phi_{,x}(x,t)\right]^2 -\lim_{x\rightarrow
-\infty}\left[\phi_{,x}(x,t)\right]^2 \right\}\nonumber\\
&&+\lim_{x\rightarrow +\infty} U\left[\phi(x,t)
\right]-\lim_{x\rightarrow -\infty} U\left[\phi(x,t) \right]\nonumber\\
&&-\beta J(t)- \left[\lim_{x\rightarrow +\infty} \phi(x,t)-
\lim_{x\rightarrow -\infty} \phi(x,t)\right]f(t).\nonumber\\
\end{eqnarray}
Differentiating Eq.~(\ref{BC1}) with respect to $t$, it is easy to see
that the first term between brace brackets on the right-hand side of
Eq.~(\ref{eq1}) is equal to zero. From Eq.~(\ref{BC2}), it follows that
the second term between brace brackets on the right-hand side of
Eq.~(\ref{eq1}) is also equal to zero. The two terms of Eq.~(\ref{eq1})
containing $U\left[\phi(x,t) \right]$ also cancel each other due to the
boundary condition (\ref{BC1}) and the periodicity of $U(z)$. Thus,
from Eq.~(\ref{BC1}) we finally obtain
\begin{equation}
\label{mainresult}
\dot{J}(t)=-\beta J(t)-\lambda Q f(t).
\end{equation}
It is important to emphasize that Eq.~(\ref{mainresult}) is a direct
 consequence of Eq.~(\ref{KG}) and the boundary conditions (\ref{BC1})
 and (\ref{BC2}). Therefore, it is an {\em exact} result valid for any
 periodic potential $U(z)$ of the type described in the paragraph below
 Eq.~(\ref{KG}), and any external field $f(t)$. Equation
 (\ref{mainresult}) appears in Ref.~\cite{Willis} as an {\em
 approximate} result obtained after neglecting the dressing due to
 phonons.

The general solution of Eq.~(\ref{mainresult}) is
\begin{equation}
\label{sol1}
J(t)=J(0) e^{-\beta t}-\lambda Q \int_{0}^{t} d t^{\prime} \;e^{-\beta
(t-t^{\prime})} f(t^{\prime}),
\end{equation}
and making use of the definition of the time-averaged energy current in
Eq.~(\ref{momentum2}), it results
\begin{eqnarray}
\label{av}
\langle J \rangle &=& \lim_{\tau\rightarrow
+\infty}\Bigg\{\frac{J(0)}{\beta \tau} (1-e^{-\beta \tau}) -\frac{\lambda
Q}{\beta \tau}\int_{0}^{\tau} dt \; f(t) \nonumber \\ &&+\frac{\lambda
Q}{\beta \tau} \int_{0}^{\tau} dt \; f(t) e^{-\beta (\tau-t)} \Bigg\}.
\end{eqnarray}
The first limit in the above expression is obviously equal to zero. The
second one is also equal to zero as the external field is periodic with
zero time-average. The integral appearing in Eq.~(\ref{av}) can be
bounded using the fact that
\begin{eqnarray}
\left|\int_{0}^{\tau} dt \; f(t)
e^{-\beta (\tau-t)}\right|&\leq& \int_{0}^{\tau} dt \; \left|f(t)\right|
e^{-\beta (\tau-t)}\nonumber\\
&\leq& \frac{f_m}{\beta} \left(1-e^{-\beta \tau}\right),
\end{eqnarray}
where $f_m$ is an upper bound of $|f(t)|$ and, thus, the third limit in
Eq.~(\ref{av}) is also equal to zero. {\em We conclude that $\langle
J\rangle=0$ for any periodic potential $U(z)$ of the type described in
the paragraph below Eq.~(\ref{KG}), and any bounded, zero time-averaged
periodic field $f(t)$.}

\begin{figure}
\includegraphics[width=6.5cm,height=8.0cm,angle=-90]{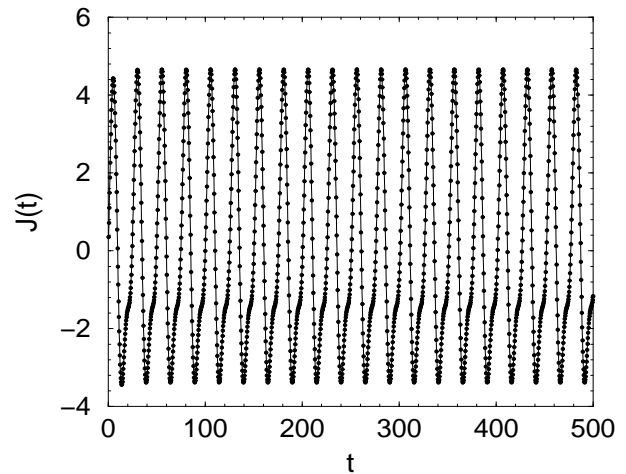}
\caption{\label{fig1} Evolution of the energy current $J(t)$
  corresponding to the sine-Gordon equation with the biharmonic field in
  Eq.~(\ref{force}) and the initial conditions corresponding to a static
  kink-like solution of the unperturbed sine-Gordon equation [i.e.,
  $f(t)=\beta=0$]. The parameter values are the same as in Fig.~3 of
  Ref.~\cite{Willis}: $\beta=0.15$, $\epsilon_{1}=0.16$,
  $\epsilon_{2}=\epsilon_{1}/\sqrt{2}$, $\omega=0.25$, and
  $\theta=1.61-\pi$. The points represents the results obtained by
  solving numerically the sine-Gordon equation and the solid line the
  analytical expression in Eq.~(\ref{sol2}). The agreement is obviously
  excellent since Eq.~(\ref{sol2}) is exact.  The time-averaged energy
  current $\langle J\rangle$ is zero.}
\end{figure}

To corroborate this result, we have solved numerically the sine-Gordon
equation for the case of the biharmonic field in Eq.~(\ref{force}). We
have considered the initial conditions for $\phi(x,0)$ and
$\phi_{,t}(x,0)$ corresponding to a static kink-like solution of the
unperturbed sine-Gordon equation [i.e., $f(t)=\beta=0$], so that,
$\lambda=2 \pi$, $Q=1$, and $J(0)=0$. For this particular choice, the
energy current in Eq.~(\ref{sol1}) reads
\begin{eqnarray}
J(t)&=&\frac{2 \pi  \epsilon_1}{\beta^2+\omega^2} \left[ -\beta e^{-\beta
t}+\beta \cos (\omega t)+\omega\sin (\omega t)\right] \nonumber \\ &&
 +\frac{2 \pi \epsilon_2}{\beta^2+4\omega^2} [ -e^{-\beta t} (\beta \cos
\theta +2 \omega \sin \theta)\nonumber \\&& +\beta \cos (2 \omega
t+\theta)+2\omega\sin(2 \omega t+\theta) \label{sol2}].
\end{eqnarray}
Figure~\ref{fig1} shows the perfect agreement between the above
expression and the numerical simulation of the sine-Gordon
equation. This agreement is not surprising at all since Eq.~(\ref{sol2})
is exact. The time-averaged energy current $\langle J\rangle$ is zero.

It is important to emphasize that the result in this comment is not
applicable when the external field not only depends on $t$ but also on
$x$. In that case Eq.~(\ref{mainresult}) cannot be obtained and, in
principle, it is possible to observe a non-vanishing time-averaged
energy current. A field of this kind has been considered in
Ref.~\cite{Flach}, where $f(x,t)=E(t)+\xi(x,t)$, with $E(t)$ being an ac
field with zero mean and $\xi(x,t)$ a Gaussian white noise.

\begin{acknowledgments}
We acknowledge financial support from the Ministerio de Ciencia y
Tecnolog\'{\i}a of Spain under the grants BFM2001-3878 (N.R.Q),
BFM2003-03015 (B.S-R), and BFM2002-03822 (J.C-P), and from la Junta de
Andaluc\'{\i}a.
\end{acknowledgments}


\begin{thebibliography}{88}
\bibitem{Flach}  S.\ Flach, Y.\ Zolotaryuk, A.\ E.\ Miroshnichenko,
and M.\ V.\ Fistul. Phys.\ Rev.\ Lett.\ {\bf 88}, 184101 (2002).

\bibitem{Salerno}  M. Salerno and Y. Zolotaryuk.
Phys.\ Rev.\ E {\bf 65}, 056603 (2002).

\bibitem{Niurka1} M. Salerno and N. R.\ Quintero,
Phys. Rev. E, {\bf 65} 025602(R) (2002); N. R.\ Quintero, B. S\'anchez-Rey and
M. Salerno, nlin.SI/0405023.

\bibitem{Costant} G.\ Costantini, F.\ Marchesoni and
M.\ Borromeo, Phys.\ Rev.\ E {\bf 65}, 051103 (2002).

\bibitem{Niurka2} L.\ Morales-Molina, N.\ R.\ Quintero,
F.\ G.\ Mertens,  and A.\ S\'anchez,
 Phys.\ Rev.\ Lett.\ {\bf 91}, 234102 (2003)

\bibitem{Willis} C. R. Willis and M. Farzaneh, Phys. Rev. E {\bf 69}, 056612
(2004).

\bibitem{Kivshar} Oleg M.\ Braun and Yuri S. Kivshar, {\it The Frenkel-Kontorova Model:
Concepts, Methods and Applications} (Springer-Verlag, Berlin-Heidelberg, 2004).

\bibitem{Ratchet}  P.\ Reimann  Phys.\ Rep.\  {\bf 361}, 57
(2002).



\end{thebibliography}
\end{document}